


\font\twelverm=cmr10  scaled 1200   \font\twelvei=cmmi10  scaled 1200
\font\twelvesy=cmsy10 scaled 1200   \font\twelveex=cmex10 scaled 1200
\font\twelvebf=cmbx10 scaled 1200   \font\twelvesl=cmsl10 scaled 1200
\font\twelvett=cmtt10 scaled 1200   \font\twelveit=cmti10 scaled 1200
\font\twelvesc=cmcsc10 scaled 1200
\skewchar\twelvei='177   \skewchar\twelvesy='60


\def\twelvepoint{\normalbaselineskip=12.4pt plus 0.1pt minus 0.1pt
  \abovedisplayskip 12.4pt plus 3pt minus 9pt
  \belowdisplayskip 12.4pt plus 3pt minus 9pt
  \abovedisplayshortskip 0pt plus 3pt
  \belowdisplayshortskip 7.2pt plus 3pt minus 4pt
  \smallskipamount=3.6pt plus1.2pt minus1.2pt
  \medskipamount=7.2pt plus2.4pt minus2.4pt
  \bigskipamount=14.4pt plus4.8pt minus4.8pt
  \def\rm{\fam0\twelverm}          \def\it{\fam\itfam\twelveit}%
  \def\sl{\fam\slfam\twelvesl}     \def\bf{\fam\bffam\twelvebf}%
  \def\mit{\fam 1}                 \def\cal{\fam 2}%
  \def\sc{\twelvesc}               \def\tt{\twelvett}
  \def\sf{\twelvesf}
  \textfont0=\twelverm   \scriptfont0=\tenrm   \scriptscriptfont0=\sevenrm
  \textfont1=\twelvei    \scriptfont1=\teni    \scriptscriptfont1=\seveni
  \textfont2=\twelvesy   \scriptfont2=\tensy   \scriptscriptfont2=\sevensy
  \textfont3=\twelveex   \scriptfont3=\twelveex  \scriptscriptfont3=\twelveex
  \textfont\itfam=\twelveit
  \textfont\slfam=\twelvesl
  \textfont\bffam=\twelvebf \scriptfont\bffam=\tenbf
  \scriptscriptfont\bffam=\sevenbf
  \normalbaselines\rm}



\def\beginlinemode{\endmode
  \begingroup\parskip=0pt \obeylines\def\\{\par}\def\endmode{\par\endgroup}}
\def\beginparmode{\endmode
  \begingroup \def\endmode{\par\endgroup}}
\let\endmode=\par
{\obeylines\gdef\
{}}
\def\singlespace{\baselineskip=\normalbaselineskip}

\def\oneandahalfspace{\baselineskip=\normalbaselineskip
  \multiply\baselineskip by 3 \divide\baselineskip by 2}
\def\doublespace{\baselineskip=\normalbaselineskip \multiply\baselineskip by 2}

\newcount\firstpageno
\firstpageno=2
\footline={\ifnum\pageno<\firstpageno{\hfil}\else{\hfil\twelverm\folio\hfil}\fi}

\def\toppageno{\global\footline={\hfil}\global\headline
  ={\ifnum\pageno<\firstpageno{\hfil}\else{\hfil\twelverm\folio\hfil}\fi}}
\let\rawfootnote=\footnote              
\def\footnote#1#2{{\rm\singlespace\parindent=0pt\parskip=0pt
  \rawfootnote{#1}{#2\hfill\vrule height 0pt depth 6pt width 0pt}}}
\def\raggedcenter{\leftskip=4em plus 12em \rightskip=\leftskip
  \parindent=0pt \parfillskip=0pt \spaceskip=.3333em \xspaceskip=.5em
  \pretolerance=9999 \tolerance=9999
  \hyphenpenalty=9999 \exhyphenpenalty=9999 }
\def\dateline{\rightline{\ifcase\month\or
  January\or February\or March\or April\or May\or June\or
  July\or August\or September\or October\or November\or December\fi
  \space\number\year}}
\def\received{\vskip 3pt plus 0.2fill
 \centerline{\sl (Received\space\ifcase\month\or
  January\or February\or March\or April\or May\or June\or
  July\or August\or September\or October\or November\or December\fi
  \qquad, \number\year)}}


\hsize=6.5truein
\hoffset=0.0truein
\vsize=8.5truein
\voffset=0.25truein
\parskip=\medskipamount
\toppageno
\twelvepoint
\doublespace
\def\\{\cr}
\overfullrule=0pt 




\def\title#1{                   
   \null \vskip 3pt plus 0.3fill \beginlinemode
   \doublespace \raggedcenter {\bf #1} \vskip 3pt plus 0.1 fill}

\def\author                     
  {\vskip 3pt plus 0.1fill \beginlinemode \doublespace \raggedcenter}

\def\affil                      
  {\vskip 4pt \beginlinemode \oneandahalfspace \raggedcenter \it}

\def\abstract                   
  {\vskip 3pt plus 0.1fill \subhead {Abstract:}
   \beginparmode \narrower \oneandahalfspace }

\def\endtopmatter               
  {\vskip 3pt plus 0.1fill \endpage \body}

\def\body                       
  {\beginparmode}               

\def\head#1{                    
   \goodbreak \vskip 0.4truein  
  {\immediate\write16{#1} \raggedcenter {\sc #1} \par}
   \nobreak \vskip 3pt \nobreak}

\def\subhead#1{                 
  \vskip 0.25truein             
  {\raggedcenter {\it #1} \par} \nobreak \vskip 3pt \nobreak}

\def\beneathrel#1\under#2{\mathrel{\mathop{#2}\limits_{#1}}}

\def\refto#1{${\,}^{#1}$}       

\newdimen\refskip \refskip=0pt
\def\references         
  {\head{References}    
   \beginparmode \frenchspacing \parindent=0pt \leftskip=\refskip
   \parskip=0pt \everypar{\hangindent=20pt\hangafter=1}}

\gdef\refis#1{\item{#1.\ }}                     

\gdef\journal#1, #2, #3 {               
    {\it #1}, {\bf #2}, #3.}            


\def\refstyleNP{                        
  \refskip=20pt
  \gdef\refto##1{ [##1]}                
  \gdef\refis##1{\item{##1)\ }}         
  \gdef\journal##1, ##2, ##3, ##4 {     
    \rm ##1 {\bf ##2}, ##3 (##4).}}



\def\endreferences{\body}

\def\figurecaptions             
  {\endpage \beginparmode \head{Figure Captions}
   \parskip=3pt \everypar{\hangindent=20pt\hangafter=1} }

\def\endpage                    
  {\vfill\eject}

\def\endpaper   {\endmode\vfill\supereject}
\def\endjnl     {\endpaper\end}


\def\ref#1{ref.{#1}}                    
\def\Ref#1{Ref.{#1}}                    
\def\[#1]{[\cite{#1}]}
\def\cite#1{{#1}}


\def\(#1){(\call{#1})}
\def\call#1{{#1}}
\def\frac#1#2{{#1 \over #2}}

\def\12{{1\over2}}

\def\sla{\raise.15ex\hbox{$/$}\kern-.57em}
\def\leaderfill{\leaders\hbox to 1em{\hss.\hss}\hfill}
\def\twiddle{\lower.9ex\rlap{$\kern-.1em\scriptstyle\sim$}}
\def\bigtwiddle{\lower1.ex\rlap{$\sim$}}
\def\gtwid{\mathrel{\raise.3ex\hbox{$>$\kern-.75em\lower1ex\hbox{$\sim$}}}}
\def\ltwid{\mathrel{\raise.3ex\hbox{$<$\kern-.75em\lower1ex\hbox{$\sim$}}}}
\def\square{\kern1pt\vbox{\hrule height 1.2pt\hbox{\vrule width 1.2pt\hskip 3pt
   \vbox{\vskip 6pt}\hskip 3pt\vrule width 0.6pt}\hrule height 0.6pt}\kern1pt}
\def\tdot#1{\mathord{\mathop{#1}\limits^{\kern2pt\ldots}}}

\def\pmb#1{\setbox0=\hbox{#1}%
  \kern-.025em\copy0\kern-\wd0
  \kern  .05em\copy0\kern-\wd0
  \kern-.025em\raise.0433em\box0 }



\refstyleNP

\catcode`@=11
\newcount\r@fcount \r@fcount=0
\newcount\r@fcurr
\immediate\newwrite\reffile
\newif\ifr@ffile\r@ffilefalse
\def\w@rnwrite#1{\ifr@ffile\immediate\write\reffile{#1}\fi\message{#1}}

\def\writer@f#1>>{}
\def\referencefile{
  \r@ffiletrue\immediate\openout\reffile=\jobname.ref%
  \def\writer@f##1>>{\ifr@ffile\immediate\write\reffile%
    {\noexpand\refis{##1} = \csname r@fnum##1\endcsname = %
     \expandafter\expandafter\expandafter\strip@t\expandafter%
     \meaning\csname r@ftext\csname r@fnum##1\endcsname\endcsname}\fi}%
  \def\strip@t##1>>{}}

\def\citeall#1{\xdef#1##1{#1{\noexpand\cite{##1}}}}
\def\cite#1{\each@rg\citer@nge{#1}}     

\def\each@rg#1#2{{\let\thecsname=#1\expandafter\first@rg#2,\end,}}
\def\first@rg#1,{\thecsname{#1}\apply@rg}       
\def\apply@rg#1,{\ifx\end#1\let\next=\relax
\else,\thecsname{#1}\let\next=\apply@rg\fi\next}

\def\citer@nge#1{\citedor@nge#1-\end-}  
\def\citer@ngeat#1\end-{#1}
\def\citedor@nge#1-#2-{\ifx\end#2\r@featspace#1 
  \else\citel@@p{#1}{#2}\citer@ngeat\fi}        
\def\citel@@p#1#2{\ifnum#1>#2{\errmessage{Reference range #1-#2\space is bad.}
    \errhelp{If you cite a series of references by the notation M-N, then M and
    N must be integers, and N must be greater than or equal to M.}}\else%
 {\count0=#1\count1=#2\advance\count1
by1\relax\expandafter\r@fcite\the\count0,%
  \loop\advance\count0 by1\relax
    \ifnum\count0<\count1,\expandafter\r@fcite\the\count0,%
  \repeat}\fi}

\def\r@featspace#1#2 {\r@fcite#1#2,}    
\def\r@fcite#1,{\ifuncit@d{#1}          
    \expandafter\gdef\csname r@ftext\number\r@fcount\endcsname%
    {\message{Reference #1 to be supplied.}\writer@f#1>>#1 to be supplied.\par
     }\fi%
  \csname r@fnum#1\endcsname}

\def\ifuncit@d#1{\expandafter\ifx\csname r@fnum#1\endcsname\relax%
\global\advance\r@fcount by1%
\expandafter\xdef\csname r@fnum#1\endcsname{\number\r@fcount}}

\let\r@fis=\refis                       
\def\refis#1#2#3\par{\ifuncit@d{#1}
    \w@rnwrite{Reference #1=\number\r@fcount\space is not cited up to now.}\fi%
  \expandafter\gdef\csname r@ftext\csname r@fnum#1\endcsname\endcsname%
  {\writer@f#1>>#2#3\par}}

\def\r@ferr{\endreferences\errmessage{I was expecting to see
\noexpand\endreferences before now;  I have inserted it here.}}
\let\r@ferences=\references
\def\references{\r@ferences\def\endmode{\r@ferr\par\endgroup}}

\let\endr@ferences=\endreferences
\def\endreferences{\r@fcurr=0
  {\loop\ifnum\r@fcurr<\r@fcount
    \advance\r@fcurr by 1\relax\expandafter\r@fis\expandafter{\number\r@fcurr}%
    \csname r@ftext\number\r@fcurr\endcsname%
  \repeat}\gdef\r@ferr{}\endr@ferences}


\let\r@fend=\endpaper\gdef\endpaper{\ifr@ffile
\immediate\write16{Cross References written on []\jobname.REF.}\fi\r@fend}

\catcode`@=12

\citeall\refto          
\citeall\ref            %
\citeall\Ref            %


\def\undertext#1{$\underline{\smash{\hbox{#1}}}$}



\title
{\undertext
{\bf
Embedded Defects
}}
\author
{Manuel $\rm Barriola^a$, Tanmay $\rm Vachaspati^b$, Martin $\rm Bucher^c$}
\affil
{
${}^a$Harvard-Smithsonian Center for Astrophysics,
60 Garden Street, Cambridge, MA 02138.
\smallskip
${}^b$Tufts Institute of Cosmology, Department of Physics and Astronomy,
Tufts University, Medford, MA 02155.
\smallskip
${}^c$School of Natural Science, Institute for Advanced Study,
Olden Lane, Princeton, New Jersey 08540.
}

\abstract

We give a prescription for embedding classical solutions and, in
particular, topological defects
in field theories which are invariant under symmetry groups that are not
necessarily simple. After providing examples of embedded defects
in field theories based on simple groups, we consider the electroweak
model and show that it contains the $Z$ string and a one parameter
family of strings called the $W(\alpha )$ string. It is argued that,
although the members of this family are gauge equivalent when
considered in isolation, each member should be considered distinct
when multi-string solutions are considered. We then turn to the issue
of stability of embedded defects and demonstrate the instability
of a large class of such solutions in the absence of bound
states or condensates. The $Z$ string is shown to be unstable when
the Weinberg angle ($\theta_w$) is $\pi /4$ for all values of the
Higgs mass. The $W$ strings are also shown to be unstable for a large
range of parameters. Embedded monopoles suffer from the Brandt-Neri-Coleman
instability. A simple physical understanding of this instability is provided
in terms of the phenomenon of W-condensation. Finally, we connect the
electroweak string solutions to the sphaleron: ``twisted'' loops of W string
and finite segments of W and Z strings collapse into the sphaleron
configuration, at least, for small values of $\theta_w$.

\endtopmatter

\head{1. INTRODUCTION}

Topological defects are classical solutions of certain field theories
and have been known for nearly three decades. These
include domain walls, strings and monopoles. However, few field theories
admit the required topology and the standard model of the electroweak
interactions\refto{swas} lacks any topological defects.
Over the last few years, it has been realized\refto{vb} that even if the
non-trivial topology required for the existence of a defect
is absent in a field theory, it may be possible to have
defect-like solutions. The idea is simply that topological defects
may be ``embedded'' in such topologically trivial field theories.
Embedded defect solutions are very common and even the electroweak
model admits string solutions. It is the properties of these solutions
that we wish to explore in this paper.

A crucial difference
between topological and embedded defects is that the stability of
the former is guaranteed by topology while the embedded defect is
generally unstable under small perturbations. Therefore, if embedded
defects are to be significant, some mechanism by which they
can be stabilized must be found. At least one embedded defect - the
semilocal string\refto{tvaa, mh} - is stable by itself and electroweak strings
can be locally stable also\refto{tv, mjlptv, wp}.
A general mechanism for stabilizing embedded defects was proposed
in Ref. \cite{tvrw}, where it was shown that scalar bound states on
electroweak strings vastly improve their stability. It was also
argued that fermionic bound states would improve the string stability
and that this mechanism of stabilizing solutions would apply to
other saddle point solutions as well. Hence, the possibility that
stable embedded defects exist in the real world must be taken seriously.

In this paper we shall investigate the existence and stability of
embedded defects with particular emphasis on defects in the
electroweak model. We shall first consider an arbitrary symmetry
breaking $G \rightarrow H$ and derive the conditions under which
embedded defects are possible (Section 2). In doing this,
we clarify the analysis in Ref. \cite{vb}
where we had only considered the
case of a simple group $G$; here we also treat the case when
$G$ is not simple. This extension has direct relevance since the
electroweak model is based on $G = SU(2) \times U(1)$ which is not
simple. We also find that a suitable choice of basis in the Lie algebra of
$G$ reduces the six conditions of Ref. \cite{vb} to two non-trivial conditions.

In Section 3, we apply the general analysis of Section 2 to a few
specific examples. These include the simplest embedded defect we could
think of - a domain wall embedded in a global $U(1)$ model; then we
construct the $O(3) \rightarrow O(2)$ string of Ref. \cite{vb} and the known
string solutions of the electroweak model. We provide further insight
into electroweak strings and show that there is a one parameter
family of string solutions - the $W(\alpha )$ string, with $\alpha$
being the parameter. All these string solutions are gauge equivalent
in isolation but should be considered as being distinct when multi-string
solutions are considered.

We turn to the issue of stability in Section 4. We first show that
embedded global defects are unstable by constructing a sequence of
field configurations that continuously lower the energy of the embedded
global defect. This very construction is then applied to embedded
{\it gauge} defects when the group $G$ is simple and we find that they
are unstable provided a certain condition on the group
generators is satisfied. This analysis immediately shows that the
$O(4) \rightarrow O(3)$ monopole\refto{vb}, the $O(3) \rightarrow
O(2)$ string and the electroweak $Z-$string at $\sin^2\theta_W = 0.5$
are all unstable.

Embedded monopoles fall into the class of ``non-topological'' monopoles
considered by Brandt and Neri\refto{rbfn} and by Coleman\refto{sc}, who
showed that such
monopoles always suffer from a long range instability. We have found
a simple physical understanding of the Brandt-Neri-Coleman instability
in terms of the phenomenon of W-condensation\refto{japo}.
This connection is described in Section 5.

It requires a little more cleverness to show that the electroweak
$W$ string is unstable. Fortunately,
our realization that the sphaleron and electroweak strings are equivalent
(discussed in Section 7)
and also the elegant derivation of the sphaleron instability can be
combined to show that the $W$ string is unstable in the absence of
bound states (Section 6).

In addition to strings, the electroweak model is known to contain a
saddle-point solution called the ``sphaleron''. The
sphaleron is an important solution
because it mediates baryon number violating processes. We discuss
the connection\refto{tvpascos} of the sphaleron to the electroweak string
in Section 7.
Our arguments show that the sphaleron can be interpreted as a
collapsed segment or loop of electroweak string.

We summarize our findings in Section 8.

\head{2. EMBEDDING SOLUTIONS}

In this section we study the conditions for the
existence of embedded defects. We are going to consider
a field theory invariant under a symmetry group $G$ that contains
$n$ simple factors, i.e.
$G = G_1 \times ...G_k \times ... G_n$. This is a generalization
of Ref. \cite{vb} in which only a simple group $G$ was considered.

The group is characterized by the Lie algebra
$$
\eqalign{
[\tau^{a_k}, \tau^{b_k}] &= f^{a_k b_k}_{\; \; \; \; \; \; \; c_k}
        \tau^{c_k}, \cr
[\tau^{a_k}, \tau^{b_{k'}}] &= 0  \qquad k \not= k' ,}
\eqno (2.1)
$$
where $\tau^{a_k}$ and $f^{a_k b_k}_{\; \; \; \; \; \; \; c_k}$ are
the generators and structure constants of the symmetry group $G_k$.
The energy functional for a static field configuration with only one
Higgs field is
$$
E = \int d^3 x \left [ {1 \over 4}Tr(G_{ij} G_{ij} )
+ [D_i \phi ] ^{\dag} [D_i \phi ] + V[\phi ] \right ]
\eqno (2.2)
$$
with
$$
G_{ij} \equiv G_{ij}^{a_k} \tau^{a_k}
\eqno (2.3)
$$
and where a sum over repeated indices is implied, $a_k$ runs over all the
generators of $G_k$ with $k=1...n$ and $i,j=1,2,3$ are spatial indices.
The Higgs potential, $V[\phi ]$, is required to lead to spontaneous symmetry
breaking but is otherwise arbitrary.

In what follows, we shall choose our group generators to satisfy
$$
Tr(\tau^{a_k} \tau^{b_l}) = \delta^{{a_k} {b_l}} \delta^{kl} \ .
\eqno (2.4)
$$
In this basis the structure constants are antisymmetric in all three
indices and the indices can be freely raised or lowered.

The covariant derivatives and field strength are defined by
$$
\eqalign{
D_i& = \partial_i + i g_k A_i^{a_k} \tau^{a_k} \cr
G_{ij} ^{a_k}&= \partial_i A_j ^{a_k} - \partial_j A_i ^{a_k} +
g_k f^{a_k b_k c_k} A_i ^{b_k} A_j ^{c_k} }
\eqno (2.5)
$$
where all indices are summed over in the covariant derivative and only
$b_k$ and $c_k$ are summed over in the field strength (no sum over $k$).
$A_i ^{a_k}$ is the gauge field associated with the generator
$\tau ^{a_k}$ and $g_k$ is the gauge coupling constant for $G_k$.

To simplify our notation we are going to define
$$
\eqalign{
T^{a_k} &= g_k \tau ^{a_k} \cr
{\cal F}^{a_k b_k c_k} &= g_k f^{a_k b_k c_k}. }
\eqno (2.6)
$$
for every $k$.
 From (2.1) the commutation relations for the generators are,
$$
[T^a , T^b] = {\cal F}^{a b c} T^c ,
\eqno (2.7)
$$
where unsubscripted Roman indices run over ${\it all}$ group generators.
With these rescalings,
$$
\eqalign{
D_i& = \partial_i + i A_i^{a} \tau^{a} \cr
G_{ij} ^{a}&= \partial_i A_j ^{a} - \partial_j A_i ^{a} +
{\cal F}^{a b c} A_i ^b A_j ^c \ .
}
\eqno (2.8)
$$

We are going to consider the possibility that the gauge fields that
describe the embedded defect may be a linear combination of the
$A_j ^{a}$. For this purpose, let us define
$$
A^a _j = \Lambda ^a _b B^b _j
\eqno (2.9)
$$
where $\Lambda$ is an orthogonal matrix ($\Lambda^T \Lambda=1$).
In terms of the new (rotated) gauge fields, the field strength is,
$$
G_{ij}^a = \Lambda ^a _b ( \partial_i B_j ^b - \partial _j B_i ^b ) +
          {\cal F}^{abc} \Lambda ^b _d \Lambda ^c _e B_i ^d B_j ^e.
\eqno (2.10)
$$
We define
$$
\Lambda ^a_b \tilde{\cal F}^{bde}={\cal F}^{abc}
\Lambda^b_d \Lambda^c_e
\eqno (2.11)
$$
so that
$$
G^a_{ij} = \Lambda^a_b \left[ \partial_i B^b_j - \partial_j B^b_i +
             {\tilde {\cal F}}^{bde} B^d_i B^e_j \right].
\eqno (2.12)
$$
Since $\Lambda$ is orthogonal, this gives,
$$
G_{ij}^a G_{ij}^a = \left[ \partial_i B_j^a -\partial_j B_i^a +
                   \tilde{\cal F}^{abc} B_i^b B_j^c \right]^2 \ .
\eqno (2.13)
$$

We next look at the kinetic term of (2.2). If we define
$$
{\cal T}^a = \Lambda^a_b T^b ,
\eqno (2.14)
$$
the covariant derivative will be
$$
D_i \phi = \partial_i + i {\cal T}^a B^a_i .
\eqno (2.15)
$$
It is easy to check that
$\tilde{{\cal F}}^{abc}$ are the structure constants for the ${\cal T}$
generators. Hence, in terms of the rescaled generators and fields, the
general energy functional in (2.2), is identical
to the energy functional
for a field theory which is invariant under transformations belonging
to a simple group. But the conditions for the existence of embedded
defects in the case of a simple symmetry group were already obtained
in Ref. \cite{vb} and so we can directly use those results. Instead of
stating those conditions, however, we will follow a more pedagogical
approach and describe how embedded solutions can be constructed and
finally state the conditions that need to be satisfied.

The {\it first} step is to look for a subgroup $G_{emb}$ of $G$.
Let the generators of $G_{emb}$ be ${\cal T}^\alpha$ where the index on
the generators of $G_{emb}$ are {\it unbarred} Greek indices.
Then, since the subalgebra $\{ {\cal T}^\alpha \}$ closes,
we have $\tilde{\cal F}^{\alpha \beta {\bar\gamma}}=0$, where {\it barred}
Greek indices denote generators other than those generating $G_{emb}$.
(Recall that we are working in the basis given by (2.4) and so
the structure constants are antisymmetric in all their indices. This
means that ${\tilde F}^{\alpha {\bar \beta} \gamma} =0 =
{\tilde F}^{{\bar \alpha} \beta \gamma}$, simply because $G_{emb}$ is a
subgroup\refto{vb}). Using (2.11) we can write these conditions
in terms of the structure constants that correspond
to the ``un-mixed'' generators,
$$
\sum_{k} g_k \Lambda^{\alpha}_{a_k} \Lambda^{\beta}_{b_k}
   \Lambda^{\bar \gamma}_{c_k} f^{a_k b_k c_k} = 0 .
\eqno (2.16)
$$
Note that, in this form, the condition depends on the coupling constants
$g_k$.

Now $G_{emb}$ acts on a subspace of the vacuum manifold and not
necessarily on the whole manifold. Then the Higgs field may be decomposed:
$\phi = \psi + \phi_{\perp}$ where $\psi$ forms a non-trivial
irreducible representation of
$G_{emb}$ and $\phi_{\perp}$ is orthogonal to all $\psi$:
$Re( \psi ^{\dag} \phi_{\perp} ) = 0$.
The generators ${\cal T} ^\alpha$ mix the components of $\psi$
but do nothing to $\phi_\perp$ while the generators
${\cal T}^{\bar \alpha}$ will
mix the components of $\phi_\perp$ and also $\psi$ and $\phi_\perp$
among themselves.
We will be interested in the case when $\phi$
acquires a vacuum expectation value such that $G_{emb}$ is spontaneously
broken down to a subgroup $H_{emb}$ and when
the symmetry breaking $G_{emb} \rightarrow H_{emb}$ contains
topological defects\footnote*{Our considerations apply equally
well to solutions that are not topological but are still present
in the embedded theory $G_{emb} \rightarrow H_{emb}$.}.
Let this topological defect solution be
denoted by $\phi_{emb}$, $[B_\mu ^\alpha ]_{emb}$. Since the topological
defect is due to the non-trivial topology associated with the breaking
of $G_{emb}$, $\phi_{emb}$ consists of a non-trivial $\psi$ and
vanishing $\phi_\perp$.
That is, $\phi_{emb} = \psi_{emb} + (\phi_{\perp})_{emb}$
where $(\phi_{\perp})_{emb} = 0$.

The {\it second} step in the construction of the embedded defect is to set up
a candidate embedded defect configuration:
$$
\phi_{emb} = \psi_{emb} \ , \ \ \
B_\mu ^\alpha =[B_\mu ^\alpha ]_{emb} \ , \ \ \
B_\mu ^{\bar \alpha} = 0
\eqno (2.17)
$$

The {\it third} and final step is to check if the candidate configuration
satisfies the constraints derived in Ref. \cite{vb}:
$$
\psi ^{\dag} {\cal T}^{\bar{\alpha}} \phi_{emb} = 0
\eqno (2.18)
$$
for all $\psi$, and
$$
{V,}_{\perp} (\phi_{emb}) = 0 \ .
\eqno (2.19)
$$
where the derivative is with respect to directions along $\phi_\perp$.
If the candidate solution does satisfy these conditions, then it is a
legitimate embedded defect.

The condition (2.18) has the interpretation that infinitesimal group elements
not belonging to $G_{emb}$ should translate points in the subspace of the
vacuum manifold covered by $\psi$ in directions that are orthogonal to
the subspace. And the condition (2.19) ensures that the potential term
in the energy functional is extremized by the candidate solution.

For completeness, we give the condition (2.18) in terms of the original
unscaled variables occurring in the energy functional (2.2):
$$
\sum _k g_k \Lambda ^{\bar \alpha} _{a_k} \tau ^{a_k} \phi_{emb} = 0 \ .
\eqno (2.20)
$$
Once again the condition depends on the gauge coupling constants.

\head{3. EXAMPLES}

We are going to describe several examples in this section that
will clarify the idea of embedded defects. We will first consider
solutions of field theories that are invariant under a simple group $G$
and then study the more complicated case of the Weinberg-Salam
model in which the group $G=SU(2)_L \times U(1)_Y$ is not simple.

{\it Walls -} The most trivial embedded solution is a domain wall
embedded in a global $G=U(1)$ model.
We express the Higgs field in terms of two real scalar fields $\phi ^a$,
$a=1,2$. A Lagrangian that is invariant under the global $U(1)$
rotation and describes static field configurations is,
$$
L= \partial _i \phi ^a \partial ^i \phi _a
              - \lambda \ \left( \phi^a \phi^a - \eta^2 \right )^2,
\eqno (3.1)
$$
with $i$ labeling the spatial coordinates.

As outlined in the previous section, the first step in constructing the
embedded domain wall solution is to identify a $Z_2$ subgroup. Let
us consider the $Z_2$ subgroup defined by the transformation:
$(\phi_1 , \phi_2 ) \rightarrow (- \phi_1 , \phi_2 )$.
Any non-zero vacuum expectation value of
$\phi_1$ will break this $Z_2$ subgroup completely and so the
embedded symmetry breaking is $Z_2 \rightarrow 1$. This symmetry
breaking has topological domain walls:
$\phi_1 = \eta {\rm tanh} ( x )$ and so this configuration for $\phi_1$
together with $\phi_2 = 0$ is our candidate solution.

The final step is to check if the conditions (2.18) and (2.19) are satisfied.
Here, since we are considering global symmetries, (2.18) is trivially
satisfied since the gauge coupling constants in (2.20) are zero. The
condition on the potential is also easily checked to be satisfied.

{\it Strings -} We will now construct a $U(1) \to 1$ string
solution\refto{aa, hnpo}
embedded in a model with $G=SO(3)$ symmetry and where the Higgs is in
the adjoint representation. The generators of $SO(3)$ are
$(T^a)_{bc}=i \epsilon ^{abc}$, with $a,b,c=1,2,3$. $T^a$ is also
the  generator of $O(2)$ rotations around the $a$-th axis
of group space. Let us consider the $O(2)$ subgroup generated
by $T^3$. This subgroup will be broken completely if either of the first
two components of $\Phi$ acquires a vacuum expectation value and hence there
will be topological string solutions in the embedded symmetry breaking.
So our candidate string configuration is:
$$
\Phi = f_{vor} (r) e^{i T^3 \theta} \pmatrix{1 \cr 0 \cr 0 }, \ \
    A_i^3=(A_i)_{vor}, \ \ A_i^{\bar \alpha} = 0, \ \ {\bar \alpha}=1,2
\eqno (3.2)
$$
where $r, \theta$ are cylindrical coordinates and the subscript $vor$
indicates the Nielsen-Olesen vortex solution. It is easily checked that
the conditions (2.18) and (2.19) are satisfied by this candidate configuration.

{\it Monopoles -} Next we consider the embedding of a `t Hooft-Polyakov
monopole in a model with $O(4)$ symmetry.
In this model the Higgs field $\Phi$ is in the adjoint representation
and has four real components. The
generators of $O(4)$ are: $(T^{\alpha})_{JK}=i \epsilon ^{\alpha JK}$,
$(T^{\bar \alpha})_{JK}={1 \over 2}(\delta ^{\bar \alpha I}
\delta^{4J}-\delta^{\bar \alpha J}
\delta ^{4I})$ where $\alpha , \bar \alpha =1,2,3$ and $I,J=1,2,3,4$.
The $T^{\alpha}$ are also the generators of an $O(3)$ subgroup and
if $\Phi$ gets a vacuum expectation value that is non-zero in the first
three components, this subgroup breaks down to $O(2)$. Since the
symmetry breaking $O(3) \rightarrow O(2)$ is known to lead to
topological magnetic monopoles, we can at once write down the
candidate embedded monopole solution:
$$
\Phi = \pmatrix{{\vec \phi}_{tP}\cr 0\cr} ,
\  A_i ^{\alpha} = [A_i ^\alpha ] _{tP} ,
\  A_i ^{\bar \alpha} = 0,
\eqno (3.3)
$$
where the subscript $tP$ indicates the `t Hooft-Polyakov solution.

The candidate configuration (3.3) can be checked to satisfy all the
necessary conditions to be an embedded magnetic monopole solution.
An important point that should be noted here is that the long range
gauge field of the embedded monopole is one of the three massless gauge
fields that remain in the symmetry breaking $O(4) \rightarrow O(3)$ and
that these three massless gauge fields transform in the adjoint
representation of $O(3)$.

{\it Electroweak strings -}
As an example of a group $G$ that is not simple we consider the
Weinberg-Salam\refto{swas, jct}  model of the electroweak interactions. The
symmetry breaking is: $SU(2)_L \times U(1)_Y \to U(1)$. This symmetry
breaking pattern is achieved in a Lagrangian of the general form
corresponding to the energy functional in (2.2) if we take the
Higgs field to be in the fundamental representation of $SU(2)$.
 From the definition (2.14), the generators are $T^a=\cos \theta_w \tau^a$,
for $a=1,2,3$ and $\tau^a$ are the $2 \times 2$
Pauli matrices, $T^4=\sin \theta_w I$ where $I$ is a two dimensional
unit matrix. The gauge field associated with these generators are
$W^a_\mu$ and  $B_\mu$ respectively.

As described in the previous section, the first step is to choose a
subgroup. We choose this to be the $U(1)$ subgroup generated by
$$
{\cal T}^3 = - \cos \theta_w T^3 + \sin \theta_w T^4
                   = diag(-\cos 2\theta_w , 1) \ .
\eqno (3.4)
$$

In order for the matrix $\Lambda$ to be orthogonal we have to choose,
$$
{\cal T}^4 = \sin \theta_w T^3 + \cos \theta_w T^4 \ .
\eqno (3.5)
$$
Then the non-trivial elements of the $\Lambda$ matrix are
$$
\Lambda^{1}_{1}=\Lambda^{2}_{2}=1,
\quad \Lambda_{3}^{3}= - \Lambda^{4}_{4} = -\cos \theta_w, \quad
     \Lambda^{3}_{4}=\Lambda^{4}_{3}=\sin \theta_w \ .
\eqno (3.6)
$$

Now the candidate embedded string solution may be written down:
$$
\phi_{emb} = f_{vor}(r) e^{i {\cal T}^3 \theta } \varphi_0 .
\eqno (3.7)
$$
where, we take,
$$
\varphi_0 = \pmatrix{0\cr 1\cr} \ .
\eqno (3.8)
$$
Here, $(r,\theta )$ are polar coordinates.

It is easy to check that the condition (2.18) for the existence
of the embedded defect is satisfied because
$$
\varphi _0 ^{\dag} {\cal T}^1 \varphi_0 =
           \varphi _0 ^{\dag} {\cal T}^2 \varphi_0 =
          \varphi _0 ^{\dag} {\cal T}^4 \varphi_0 = 0  \ .
\eqno (3.9)
$$
The potential condition can also be easily checked to be satisfied.

With the choice of (3.8) for $\varphi_0$, it is easy to see that the solution
corresponds to a string with $Z$ magnetic flux because
$Z_{\mu}$ is the gauge boson associated with the generator
${\cal T}^3$. Therefore the full embedded $Z$ string solution is:
$$
\phi_{emb} = f_{vor}(r) \pmatrix{0 \cr e^{i\theta} \cr },
\ \ \ Z_\mu=(A_i)_{vor}, \ \ \  W_i^1=W_i^2=0 = (A_i )_{em} \ .
\eqno (3.10)
$$

Different choices of the (``embedded'') subgroup leads to other string
solutions. The choice that we now consider is the subgroup sitting
entirely in the $SU(2)$ factor of the electroweak model and generated
by: ${\cal T}_\alpha \equiv sin \alpha \ \tau^1 + cos \alpha \ \tau ^2$,
where $\alpha$ is some
constant. Then the corresponding embedded string solution is:
$$
\phi_{emb} = f_{vor}(r) e^{i {\cal T}_\alpha \theta} \varphi_0 \ ,
\ \ \
sin \alpha W^1 _i + cos \alpha W^2 _i = (A_i)_{vor} \ ,
\eqno (3.11)
$$
and all other orthogonal combinations of gauge fields vanish.

The one parameter family of string solutions in (3.11) is called
the $W$ string since the flux in the string is purely in the
$SU(2)$ sector. Furthermore, by a global gauge transformation,
any single string solution in the family - that is, a string with any
value of $\alpha$ - may be transformed into the string configuration
with $\alpha = 0$. Explicitly, this gauge transformation is:
$$
\phi ' = exp\left [ - i {{(1+\tau ^3 )} \over 2}
                     \alpha \right ] \phi
\eqno (3.12)
$$
together with a corresponding transformation of the gauge fields.
This does not, however, mean that
multi-string solutions of different $\alpha$ can be gauge transformed
to another multi-string solution with all strings having the same value
of $\alpha$. The simplest way to see that $\alpha$ is a non-trivial
parameter is to consider a loop of $W$ string such that $\alpha$
runs from 0 to $2\pi$ as we go around the loop. The winding of
$\alpha$ around the loop is a discrete number and cannot be altered
by any non-singular gauge transformation. Hence, a loop with varying
$\alpha$ is not gauge equivalent to one with a constant value of
$\alpha$.

There are two other arguments that led us to the conclusion that
$\alpha$ is not a gauge artifact but a genuine label for different string
solutions. The first of these is that if we consider two $W$ strings with
$\alpha = 0$, they would combine to form a winding number 2 string
with $\alpha = 0$ and, in particular, would not annihilate each other
when brought together. On the other hand, an $\alpha = \pi$ string is
the anti- of the $\alpha = 0$ string and these would annihilate
each other if brought together. Therefore the system with two $\alpha = 0$
strings must be gauge inequivalent to the system with one $\alpha = 0$
string and another $\alpha = \pi $ string. The second
argument is that if we try and construct a straight $W$ string with
varying $\alpha$, we necessarily find that the gradients in $\alpha$
cause electromagnetic fields to emanate from the string and hence the
variations in $\alpha$ cannot be gauge artifacts.

To summarize, we have constructed the $Z$ string and a one parameter family
of distinct $W$ strings present in the Weinberg-Salam model of the
electroweak interactions.


\head{4. STABILITY}

In this section we consider the stability of embedded defects.
Although the question of which embedded solutions are stable
against small perturbations is not answered in its full generality,
a large class of embedded solutions are shown to be unstable
by explicitly indicating a particular instability. Qualitatively, this mode
can be described as a combination of a dilatation and a rotation
of the embedded solution into the trivial vacuum. The instability that we
have found only applies when the embedded gauge group $G_{emb}$ acts trivially
on the subspace spanned by $\phi _\perp$ and when the potential is of the
Mexican hat variety. One outcome is that {\it all} global embedded defects
in models with a Mexican hat potential are unstable to small perturbations.

Let us consider an embedded solution in a model with the energy functional
given in (2.2) and with the specific form of the potential:
$$
V[ \phi ] = \lambda \left ( \phi ^{\dag} \phi - \eta ^2 \right ) ^2 \ .
\eqno (4.1)
$$
The embedded defect solution is given in (2.17) and the Higgs field for
the solution will be denoted by $\phi_0$. Now consider the sequence of
configurations labeled by the parameter $\xi$:
$$
\eqalign{
\phi (x;\xi )&=
\cos \xi \ \phi _0(\cos \xi \ x)+\sin \xi \ \phi _\bot \cr
A_j (x;\xi )&=\cos \xi \ A_j (\cos \xi \ x)\cr }
\eqno (4.2)
$$
where $\phi _\bot $ is constant with
${\phi _{\bot}}^{\dag} \phi_{\bot}= \eta ^2 $ and
${\phi _{\bot}}^{\dag} \phi _0 (x)=0,$.
For $\xi = 0$, the configuration is the embedded defect solution
and for $\xi = \pi /2$, the configuration describes the vacuum.

We then have,
$$\eqalign{
\phi^ {\dag}(x;\xi ) \phi (x;\xi )&
= \cos ^2 \xi \ {\phi_0}^{\dag} \phi_0 + \sin ^2 \xi \eta ^2
\cr
G_{ij}^a (x;\xi ) G_{ij}^a (x;\xi )
&= \cos ^4 \xi \  G_{ij}^a(x) G_{ij}^a(x),
\cr
D_i \phi  (x;\xi )= \cos ^2 \xi &\left [\partial _i +
iA_i^{\alpha}(x){\cal T}^{\alpha} \right] \phi_0(x)
+i \cos \xi \sin \xi A_i^{\alpha}(x) {\cal T}^{\alpha} \phi_{\bot},
\cr
V[\phi (x, \xi )] &= \cos ^4 \xi \  V[\phi _0(x)]
\cr
}
\eqno (4.3)
$$
If the orthogonality condition
$$
{\cal T}^\alpha  \phi_{\bot} = 0
\eqno (4.4)
$$
is satisfied, we get
$$
[D_i \phi (x;\xi )]^\dagger  \ [D_i \phi (x;\xi )]
=\cos ^4\xi \ (D_i \phi _0 (x))^\dagger  (D_i \phi _0 (x)) \ .
\eqno (4.5)
$$
and hence, the total energy of the configuration is
$$
E(\xi )=\cos ^{d}\xi \ E(\xi =0),
\eqno (4.6)
$$
where $d$ is the dimension of the world-surface of the defect.
($d=1, 2, 3$ for monopoles, strings, and walls respectively.)
For $d > 0$ this shows that the solution is quadratically unstable
because $\xi $ paramterizes a smooth sequence of configurations
with monotonically decreasing energy
starting at the embedded defect solution ($\xi = 0$) and ending at
the vacuum ($\xi = \pi /2$).

We are now prepared to consider some specific examples. For a global
embedded defect the condition (4.4) is trivially satisfied because there are
no gauge fields. (Recall that the generators ${\cal T}^\alpha$ have
been rescaled with the gauge coupling constants as in (2.6) and so,
in the global case, ${\cal T}^\alpha = 0$.) Therefore, {\it all} embedded
global defects are unstable.

At this juncture we should point out that the existence of a sequence of
lower energy configurations such as that given in (4.6) does not say
anything about the time scales associated with the dynamical
instability\refto{kbmb}. This is because
the inertia associated with the instability could be large and this
would cause the instability to be slower. The issue is even more tricky
when we consider global defects since the energy of global strings and
global monopoles diverges and there is a possibility that the inertia
associated with the above instability would also diverge. In this case,
although (4.6) is valid, the time scale associated with the instability
is infinite and the defect would not decay. The best way to see that
this is not the case is to consider perturbations that are truncated
beyond a certain distance. For example, we could consider $\xi$ to have
the following form:
$$
\eqalign{
\xi (r < R , t) &= \xi_0 (t) \cr
\xi (r > R , t) &= \xi_0 (t)
         exp \left [ - {(r- R)} \over l \right ]  \cr
}
\eqno (4.7)
$$
where, $R$ and $l$ are some length scales that are large compared to
the core of the defect. We have reconstructed the energy of such
configurations for small values of $\xi_0$
(without the rescaling of the coordinate in (4.2))
and kept the kinetic term due to the time derivative of $\xi_0 (t)$.
This calculation confirms that the inertia of the truncated
perturbation (4.7) is not infinite and that the instability is
on a finite time scale.

We next consider the embedded gauge monopole solution constructed
in Sec. 3 and given by (3.3). The fourth component
of the vector $\Phi$ is annihilated by the action of
$T^\alpha$ the generators of $G_{emb} = O(3)$. Hence, (4.4) shows that
this embedded monopole is unstable.

We now consider two important cases for which the procedure for
demonstrating instability described above fails.
Let us first consider the $W$ string in the electroweak model with
$G=SU(2)_{L}\times U(1)_{Y}$. Here the condition (4.4)
cannot be satisfied because $U(1)_{Y}$ does
not annihilate any nonvanishing Higgs field and so the instability
does not apply.

For the embedded $Z$-string solutions in the electroweak model
the situation is slightly more complicated. It turns out that the above
argument for instability fails except for the special case $\theta _W= \pi /4$.
The string is generated by  ${\cal T}^3 = diag(-\cos 2\theta_w , 1)$
 From the orthogonality condition $Re( \phi_\bot ^{\dag} \phi _0(x)) =0$,
we can choose $\phi _\bot ^{\dag} = (1, 0) $. Therefore, the condition
for instability, eqn. (4.4), becomes $cos 2\theta _W=0$, that is,
$\theta _W=\pi /4$.

This last result is of some importance because the stability
analysis of Ref. \cite{mjlptv}
did not consider the case of very low Higgs masses and it was not clear
from the given stability diagram if it would be possible to find some
value of the Higgs mass for which the $Z$ string in the standard model
with $sin^2 \theta _w = 0.23$ could be stable. Our result here shows
that the $Z$ string is unstable for {\it all} Higgs masses at
$sin^2 \theta _w = 0.5$, making it extremely unlikely for there to be
stable solutions for smaller values of $sin^2 \theta _w$.

So far we have ignored the possibility that there may be bound states on
the embedded defects. It has been shown\refto{tvrw} that such bound states
can considerably enhance the stability of the defect. The physical reason
behind this enhancement is the same as the reason behind
the existence of non-topological solitons\refto{tl} and is discussed in
some detail in Ref. \cite{tvrw}. Mathematically, the
introduction of a bound state would result in the presence of additional
terms in the varied energy functional (4.6) that would be proportional
to $sin^2 \xi$. With these additional terms, it is possible that $\xi = 0$
describes a local minimum of the energy and so there is no instability
towards increasing $\xi$.

\head{5. INSTABILITY OF EMBEDDED MONOPOLES}

The stability of monopoles has been
studied by Brandt and Neri\refto{rbfn} and by Coleman\refto{sc}. They
find that the asymptotic magnetic field of the monopole
has an instability unless
the monopole is topological. Here we will give a simple explaination of the
Brandt-Neri-Coleman instability in the context of embedded
monopoles\refto{jp}.

The key to understanding the instability is already present in the work
on W-condensation\refto{japo}. The idea
is that a spin $s$ particle in a uniform magnetic field $\vec B = B \hat z$
has energy levels given by:
$$
E_n ^2  = (2n+1) e B - 2 e B s + k_3 ^2 + m^2
\eqno (5.1)
$$
where the mass of the particle is $m$, the electric charge is $e$
and the momentum in the $z$ direction is ${\vec k}_3$.
The first term in (5.1) is the Landau level term and is due to the
orbital motion of the particle, the second term is the spin-magnetic
field coupling, the third term is the kinetic energy due to the motion
along the magnetic field and the last term is due to the rest mass of the
particle. Note that the g-factor, that is, the numerical factor in the
spin-magnetic field coupling, has been taken to be 2. For our purpose,
since the magnetic field and the spin 1 field all belong to the adjoint
representation of a non-Abelian group, this will be true\refto{japo}.
The crucial observation is that for $s = 1$ and for large enough magnetic
fields, $E_{n=0} ^2$ can be negative and the system can decay to a state
of lower energy by the creation of spin 1 particles.
The critical magnetic field strength needed for the instability is:
$$
B_{c} = {{m^2} \over e }
\eqno (5.2)
$$

Now consider the embedded monopole discussed in Sec. 3. The magnetic field
of the monopole is one of the three gauge fields of the $O(3)$ residual
symmetry group. Hence, far away from the monopole, the effective Lagrangian
for the gauge fields is:
$$
L = -{1 \over 4} F_{\mu \nu}^a F^{\mu \nu a}
\eqno (5.3)
$$
where $a = 1,2,3$. In the asymptotic region with, say, $z > 0$,
the monopole magnetic field is simply like a uniform magnetic field.
Therefore, in this region, the field configuration is:
$$
F_{12}^{a=3} = B(z)
\eqno (5.4)
$$
and the $a = 1,2$ components are zero.
But now we can apply the W-condensation arguments to this field since
there are other spin 1 gauge fields present in the system - namely, the two
other gauge fields of the O(3) theory. These gauge fields are all massless
and hence the critical field needed for an instability is zero (eqn. (5.2)).
This means that the magnetic field of the embedded monopole suffers from the
same instability that leads to W-condensation.

The literature on W-condensation makes a further point that the phenomenon
of W-condensation actually {\it anti-screens} the applied magnetic field.
Naively, this would imply that the magnetic field of the monopole would
increase due to the instability. However, this is not true since, as pointed
out by Ambjorn and Olesen\refto{japo}, only U(1) fields are anti-screened.
The non-Abelian gauge fields become larger due to the instability but
in such a way that the field strengths become smaller. This can also
be checked to be the case for the embedded monopole and hence, the magnetic
field of the embedded monopole is diminished due to the instability.

The instability of the core structure of global defects described in
Sec. 4 also applies to the embedded monopole. Therefore the embedded
monopole suffers from {\it two} instabilities: the first is that the core
spreads out and the second is that the long range magnetic field gets
screened due to W-condensation. By considering the stability of embedded
monopoles in the presence of bound states, it seems to us that
the first of these instabilities can be avoided but that the second
instability is incurable.

\head{6. W STRING INSTABILITY}

Here we will show that the bare $W$ string is unstable for a large range of
parameters. The idea of the proof follows from the observation that
the sphaleron and the $W$ string are closely related (see the following
section) and so the instability of the sphaleron found by
Manton\refto{nm} might well apply to the $W$ string also.

The energy functional we want to consider is that given in (2.2)
with $SU(2) \times U(1)$ symmetry and the Higgs in the fundamental
representation of $SU(2)$. The potential is taken to be the Mexican
hat potential given in (4.4) together with $\eta ^2 = 1$. (This amounts
to suitably rescaling the fields and the coordinates.) The $W$ string
solution is given
in (3.11) and, to be specific, we consider the $\alpha = \pi /2$ solution.

Now consider the family of Higgs field configurations labeled by the
index $\mu$:
$$
\Phi (\mu , r, \theta , z) =
            (1-f_{vor} (r)) \pmatrix{0\cr e^{-i\mu} cos\mu \cr} +
  f_{vor} (r) \pmatrix{ sin\mu \ sin\theta \cr
                      e^{-i\mu} (cos\mu + i sin\mu \  cos\theta )\cr }
\eqno (6.1)
$$
and, the family of gauge field configurations:
$$
A_j (\mu , r, \theta ,z) = - i v_{vor} (r) ( \partial_j U ) U^{-1}
\eqno (6.2)
$$
where
$$
U = \pmatrix{e^{i\mu}(cos\mu - i sin\mu \ cos\theta )&sin\mu \ sin\theta\cr
      - sin\mu \ sin\theta&e^{-i\mu}(cos\mu +i sin\mu \ cos\theta )\cr} \ .
\eqno (6.3)
$$
Note that this family of configurations is almost identical to the
family considered by Manton for the case of the sphaleron. The
differences are that we are working in cylindrical coordinates
and that we have discarded the $\phi$ (the spherical azimuthal angle)
dependence that is present in the sphaleron.

The configurations in (6.1)-(6.3) yield the $W$ string when
$\mu = \pi /2$ and at $\mu = 0$ or $\mu = \pi$, the configuration
is that of the trivial vacuum. The parameter $\mu$ parametrizes a path
from the $W$ string configuration to the trivial vacuum.

The energy of the configurations can be found by inserting
(6.1)-(6.3) into the energy functional (2.2).
Then, after some algebra we get,
$$
\eqalign{
E(\mu ) = 2\pi \int dz dr & r \biggl [
           sin^2 \mu \left ( {{v'} \over {gr}} \right ) ^2 +
           sin^2 \mu {f'}^2 +
\cr
     &{{sin^2 \mu} \over {r^2}} [ cos^2 \mu \left \{
       v^2 (1-f)^2 - 2 v (1-f) f (1-v)       \right \} + f^2 (1-v)^2
                                ] +
\cr
     \qquad & \sin^4 \mu \ \lambda  (1 - f^2 )^2
                            \biggr ]}
\eqno (6.4)
$$
where the subscript $vor$ on the functions $f$ and $v$ has been dropped for
convenience. It is clear that $E$ is an increasing function of
$sin^2 \mu$ at $\mu = \pi /2$ provided
$$
E_1 \equiv sin^2 \mu \ cos^2 \mu \int {{dr} \over r} [
          v^2 (1-f)^2 - 2 v (1-f) (1-v)] +
      sin^2 \mu \int {{dr} \over r} f^2 (1-v)^2
\eqno (6.5)
$$
is a monotonically increasing function of $sin^2 \mu$ at $\mu = \pi /2$.
Let us denote
the two integrals in (6.5) by $I_1$ and $I_2$ respectively and write
$\xi \equiv sin^2 \mu \in [0,1]$.
By differentiating (6.5) with respect to $\xi$, we find that $E_1$
is a monotonically increasing function of $\xi$ if $I_2 > (2 \xi - 1) I_1$.
We have checked this condition numerically at $\xi = 1$
for certain values of the parameters ($\lambda$ and $g$)
and always found it to be satsified. This shows that the
$W$ string is unstable for the parameters we considered.

We wish to remark that if we could show that
$$
s(\xi ) \equiv \xi (1 - \xi ) [ v^2 (1-f)^2 - 2vf(1-f)(1-v)] +
               \xi f^2 (1-v)^2
\eqno (6.6)
$$
is an increasing function of $\xi$, then the condition regarding
$E_1$ would also be satisfied. Now it is straightforward to
show that $s(\xi )$ is maximum at $\xi = 1$ if
$$
(1+\sqrt{2} ) f (1-v) \ge v (1-f)
\eqno (6.7)
$$
for all $r$. Numerical evaluations of the Nielsen-Olesen vortex
profile have shown that (6.7) is satisfied for almost
all $r$ for a large range of values of $\lambda$.
The inequality (6.7) is violated only in the
large $r$ region, where the integrands in (6.5) are exponentially
small. So the contributions that could change the monotonic increase
of $E_1$ with $\xi$ are exponentially suppressed and $E_1 (\xi )$
is an increasing function of $\xi$ for a wide range of parameters
\footnote*{Note that a condition similar to (6.7) is assumed to hold
in the case of the sphaleron\refto{nm}.}.

\head{7. ELECTROWEAK STRINGS AND THE SPHALERON}

In this section, we connect\refto{tvpascos} the electroweak string
solutions with
the sphaleron solution discovered by Manton\refto{nm}.
In the limit $sin^2 \theta_W \rightarrow 0$, the sphaleron
solution is\refto{fknm}:
$$
\Phi = f_s (r) \pmatrix{ e^{i\phi} sin\theta \cr cos\theta\cr }
\equiv f_s (r) U \pmatrix{0\cr 1\cr}, \ \ \
A_\mu \equiv W_\mu ^a \tau^a = -i v_s (r) (\partial_\mu U ) U^{-1}
\eqno (7.1)
$$
where we are now using spherical coordinates,
$\tau^a$ are the Pauli spin matrices and the subscript {\it s} on
the radial functions $f$ and $v$ denote that these functions
are particular to the sphaleron.

The sphaleron solution in (7.1) necessarily has an accompanying
electromagnetic field which can be calculated\refto{fknm} to first order
in $\theta_W$. Alternately, it can be derived from the gauge invariant
definition of the electromagnetic field strength in the
electroweak model\refto{tvmagn}:
$$
F_{\mu \nu} ^{em} = \partial _\mu A_\nu ^{em}
                        - \partial _\nu A_\mu ^{em}
   - i4 g^{-1} \eta^{-2} sin\theta _W [ \partial _\mu \phi ^{\dag}
      \partial_\nu \phi - \partial_\nu \phi ^{\dag} \partial_\mu \phi ]
\eqno (7.2)
$$
where, the electromagnetic gauge potential is
$$
A _\mu ^{em} = sin\theta _W n^a W_\mu ^a + cos\theta _W B_\mu
\eqno (7.3)
$$
with $n^a \equiv -2\phi ^{\dag} \tau ^a \phi /\eta^2 $.

The back-reaction of the electromagnetic
field on the sphaleron configuration is second order in $\theta_W$ and
can be ignored in the limit that we are considering.

Now the matrix $U$ in (7.1) is a unitary matrix and may be written as:
$$
U = exp[ i {\hat m} \cdot {\vec \tau} \theta ]
\eqno (7.4)
$$
where, ${\hat m} = (sin\phi , cos\phi , 0)$ and $(\theta , \phi )$
are spherical angles. A comparison of (7.4) with (3.11) immediately
suggests that the sphaleron configuration (7.1) is precisely that
of a (``twisted'') loop of $W$ string in which $\alpha = \phi$.
One difference is that $\theta$ is a spherical angle and ranges
from 0 to $\pi$ in (7.4) whereas in the $W$ string it is a
cylindrical angle and ranges from 0 to $2\pi$. Another difference
is that the Higgs field vanishes only at one point in the sphaleron
while in the case of the $W$ string loop, it vanishes along a one-dimensional
closed curve. Both these differences can be reconciled by imagining a twisted
loop of $W$ string that has collapsed to a single point. In this case,
one should indeed restrict $\theta$ to go from 0 to $\pi$ and have
a vanishing Higgs field at only one point.

One could also get different interpretations of the sphaleron in terms
of electroweak strings by considering various slices of (7.1). For
example, the $xz$ and $yz$ slices of the sphaleron yield the $W$ string
for $\alpha = \pi /2$ and $\alpha = 0$ respectively. Therefore,
``stretching'' deformations of the sphaleron along any axis in the $xy$
plane would yield finite segments of $W$ strings.
Reversing this argument tells us that a segment of $W$ string (for any
$\alpha$) collapses into a sphaleron.

The final interpretation of the sphaleron that we point out seems
like the most interesting to us. It is possible to arrive at this
interpretation in two ways which we now describe.

If we look at the $xy-$slice of the sphaleron, that is, if we set
$\theta = \pi /2$ in (7.1), we find that the resulting configuration is
precisely that of the $Z$ string in (3.10) upto the profile functions
and a global gauge transformation.
Hence, if we were to stretch the sphaleron along the $z-$axis, we would get
a segment of $Z$ string. Or in other words, a finite segment of $Z$ string
collapses into a sphaleron. But we know from the work of Nambu\refto{yn}
that the  $Z$ string ends on magnetic monopoles. Hence, the sphaleron must
be  equivalent to a monopole sitting adjacent to an antimonopole along the
$z$ axis.

This interpretation can be arrived at directly by looking at the Higgs
field configuration of the electroweak monopole found by Nambu\refto{yn}:
$$
\Phi = f_m (r) \pmatrix{ e^{i\phi} sin\theta /2\cr cos\theta /2\cr }
\eqno (7.5)
$$
and comparing to the sphaleron Higgs field configuration given in
(7.1). (The gauge fields for the monopole are given by the
same formula as in (7.1).)
The two configurations are identical upto the profile functions
and, more importantly, upto a factor of 2 wherever $\theta$ appears.
Hence, as $\theta$ is varied from 0 to $\pi /2$ in the sphaleron
configuration, the full monopole configuration is mapped out. Then
as $\theta$ is varied from $\pi /2$ to $\pi$ in the sphaleron, an
antimonopole configuration is mapped out. This directly confirms that
the sphaleron can be viewed as a monopole and an antimonopole sitting
adjacent to each other.

The above interpretations lead to a picture for the space of
configurations in the vicinity of the sphaleron in the electroweak model.
The sphaleron is a saddle point solution to the electroweak equations
of motion and has one unstable mode\refto{ly}. Therefore it is
useful to think of an ordinary 2-dimensional saddle embedded in
3-dimensional space with the saddle point being the sphaleron
Now, when we deform the sphaleron to get segments of string, we are
going to higher energy configurations and so we are
climbing up the saddle. Of course, there are many ways of going to
higher energy configurations but there is a special one -
the one which goes along the ridge of the saddle. Furthermore,
once we have found this configuration, it will have two
unstable modes: the first one causes the configuration to roll down
into the sphaleron while the second instability is towards rolling off
in a direction orthogonal to the ridge. These instabilities are
exactly what we see for a finite segment of $Z$ string when
$sin^2 \theta_W$ is not too close to 1. The segment of string is
dynamically unstable to collapsing to shorter lengths and this
corresponds to rolling down the ridge into the sphaleron configuration.
The infinite $Z$ string also has an unstable mode and this can
be viewed as the mode that causes the string to slide off the ridge.

It is interesting to consider what might happen when
$sin^2 \theta_W$ is close to 1. Then the infinite $Z$ string is metastable
and the mode that corresponds to sliding off the ridge is absent. This
means that the saddle is not of the usual kind - it contains {\it two}
ridges that go up from the saddle point and the $Z$ string lies in
the valley between the two ridges. Assuming that the connection of the
sphaleron and $Z$ string remains valid at large $\theta_w$ also,
a finite segment of string will collapse into a sphaleron configuration
by rolling in the valley between the two ridges.

The sphaleron is known to be the intermediate point in processes
that violate baryon number. In the language of the saddle, if we
consider a sequence of configurations that pass from one side of
the saddle to the other side, the baryon number of the configuration
changes\refto{nm}. However, in going from one side to the
other, it is necessary to pass through either the sphaleron configuration -
which would be the least energetically expensive way - or to pass
through a string configuration. Therefore, electroweak strings will
play the same role as the sphaleron in baryon number violating processes.

At low energies, it is unfavourable to have long strings
and the shortest string - the sphaleron - will dominate all baryon
number violating processes. However, the presence of electroweak
strings can become interesting if there is a mechanism that can
stabilize string segments for a sufficiently long duration\refto{tvrw}. In
the past, mechanisms have been found that prevent topological superconducting
loops\refto{bc, rdps, ecmhnt, chhhmt, abtpds}
from collapsing and it is an open question
as to whether those mechanisms will apply in the electroweak case too.

An issue that we have not investigated but feel could be very interesting
is the possible connection of electroweak
strings with the (deformed) sphaleron solutions found by Yaffe\refto{ly}
for large values of the Higgs mass.

\head{8. SUMMARY}

We itemize our main results:

\item{(i)} We have described a procedure by which embedded defect
solutions may be constructed in Section 2. The procedure applies
whether the symmetry group of the theory is simple or not. Examples
were provided in Section 3. In particular, it was shown that the
electroweak model contains the $Z$ string and a one parameter family
of $W$ strings.

\item{(ii)} In Section 4 we have considered the stability of
embedded defects in the absence of bound states. By considering a
specific perturbation of the embedded defect solution, it was shown that
embedded global defects are unstable and that embedded gauge defects are
unstable if condition (4.4) is satisfied. By an application of this
condition, we showed that the electroweak $Z$ string is unstable when
$sin^2 \theta_W = 0.5$. By considering a separate argument (Section 6),
we showed that the $W$ string is unstable for a wide range of parameters.

\item{(iii)} In Section 5 we explained the Brandt-Neri-Coleman instability
for embedded gauge monopoles in terms of the phenomenon of W-condensation.

\item{(iv)} In Section 7, we showed that the sphaleron may be
reinterpreted as segments or loops of electroweak strings.

\head{\it{Acknowledgements:}}

We would like to thank Steve Hsu for discussions regarding baryon
number violation, Bharat Ratra for bringing Ref. \cite{ly} to our
attention and Rick Watkins for invaluable numerical help. In addition,
we would like to acknowledge useful comments by Sidney Coleman, Jaume
Garriga and Alex Vilenkin.
This work was supported in part by the National Science Foundation.

\vfill
\eject

\references

\refis{swas} S. Weinberg, Phys. Rev. Lett. {\bf 19}, 1264 (1967);
A. Salam in ``Elementary Particle Theory'', ed. N. Svarthholm, Stockholm:
Almqvist, Forlag AB, pg 367.

\refis{tvaa} T. Vachaspati and A. Ach\'ucarro, Phys. Rev. D {\bf 44},
3067 (1991).

\refis{jct} J. C. Taylor, ``Gauge Theories of Weak Interactions'',
Cambridge University Press, 1976.

\refis{hnpo} H. B. Nielsen and P. Olesen, Nucl. Phys. B{\bf{61}}, 45 (1973).

\refis{mh} M. Hindmarsh, Phys. Rev. Lett. {\bf 68}, 1263 (1992); Nucl. Phys.
{\bf B392}, 461 (1993).

\refis{jp} J. Preskill, Phys. Rev. {\bf D46}, 4218 (1992).

\refis{ecmhnt} E. Copeland, M. Hindmarsh and N. Turok,
Phys. Rev. Lett. {\bf 58}, 1910 (1987);
E. Copeland, D. Haws, M. Hindmarsh and N. Turok,
Nucl. Phys. B{\bf 306}, 908 (1988);
D. Haws, M. Hindmarsh and N. Turok, Phys. Lett. B{\bf 209},
255 (1988).

\refis{chhhmt} C. Hill, H. Hodges and M. Turner, Phys. Rev. D{\bf 37},
263 (1988).

\refis{abtpds} A. Babul, T. Piran and D. N. Spergel, Phys. Lett. B{\bf 202},
307 (1988).

\refis{rdps} R. L. Davis and E. P. S. Shellard, Nucl. Phys. B{\bf 323},
189 (1989); R. L. Davis, Phys. Lett. B{\bf 207}, 404 (1988); Phys.
Lett. B{\bf 209}, 485 (1988); Phys. Rev. D{\bf 38}, 3722 (1980).

\refis{nm} N. S. Manton, Phys. Rev. D{\bf 28}, 2019 (1983).

\refis{yn} Y. Nambu, Nucl. Phys. B{\bf 130}, 505 (1977).

\refis{bc} B. Carter, Ann. N. Y. Acad. Sci., {\bf 647}, 758 (1991).

\refis{kbmb} For a treatment of the dynamical time-scales associated with
semilocal strings, see K. Benson and M. Bucher, Nucl. Phys. {\bf B}, to be
published.

\refis{tvmagn} T. Vachaspati, Phys. Lett. {\bf B265}, 258 (1991).

\refis{japo} For a recent review, see J. Ambjorn and P. Olesen,
NBI-HE-93-17 (1993).

\refis{vb} T. Vachaspati and M. Barriola, Phys. Rev. Lett. {\bf 69},
1867 (1992).

\refis{aa} A. Abrikosov, Soviet Phys. JETP {\bf 5}, 1174 (1957).

\refis{tv} T. Vachaspati, Phys. Rev. Lett. {\bf 68}, 1977 (1992);
{\bf 69}, 216(E) (1992); Nucl. Phys. {\bf B397}, 648 (1993).

\refis{wp} W. Perkins, University of Sussex preprint, 1992.

\refis{tvrw} T. Vachaspati and R. Watkins, TUTP-92-10.

\refis{mjlptv} M. James, L. Perivolaropoulos and T. Vachaspati, Phys.
Rev. D{\bf 46} (1992) R5232; Nucl. Phys. {\bf B395}, 534 (1993).

\refis{tl} T. D. Lee, {\it In} Proceedings of the Conference on Extended
Systems in Field Theory [Phys. Rep. {\bf 23C}, 254] (1976).

\refis{fknm} F. R. Klinkhammer and N. S. Manton, Phys. Rev. D{\bf 30}, (1984)
2212.

\refis{rbfn} R. A. Brandt and F. Neri, Nucl. Phys. {\bf B161}, 253 (1979).

\refis{sc}  S. Coleman, ``The Magnetic Monopole: 50 Years Later'',
Erice Lectures (1981).

\refis{ly} L. G. Yaffe, Phys. Rev. {\bf D40}, 3463 (1989).

\refis{tvpascos} T. Vachaspati, Proceedings of the Texas/Pascos 1992
Conference, Berkeley.

\endreferences



\endjnl
\end